\documentclass{emulateapj}
\usepackage{amsmath, subfigure}
\newcommand{\colwidth}{1\linewidth}

\newcommand{\Msol}{M_{\odot}}
\newcommand{\E}[1]{\ensuremath{\times10^{#1}}}

\slugcomment{Draft Version \today.}
\shorttitle{1--5~Hz flaring in SAX J1808.4--3658}
\shortauthors{Bult et al.}

\begin{document}

\title{Discovery of 1--5~Hz flaring at high luminosity in SAX J1808.4--3658}
\author{Peter Bult}
\affil{Anton Pannekoek Institute, University of Amsterdam, Postbus 94249, 1090 GE Amsterdam, The Netherlands}
\email{p.m.bult@uva.nl}
\and
\author{Michiel van der Klis}
\affil{Anton Pannekoek Institute, University of Amsterdam, Postbus 94249, 1090 GE Amsterdam, The Netherlands}

\begin{abstract}
We report the discovery of a 1--5~Hz X-ray flaring phenomenon observed
at $>30$~mCrab near peak luminosity in the 2008 and 2011 outbursts of the accreting
millisecond X-ray pulsar SAX J1808.4--3658 in observations with the
\textit{Rossi X-ray Timing Explorer}. In each of the two outbursts this high
luminosity flaring is seen for $\sim$3 continuous days and switches on and off 
on a timescale of 1--2~hr. 

The flaring can be seen directly in the light curve, where it shows sharp spikes 
of emission at quasi-regular separation. In the power spectrum it produces a broad 
noise component, which peaks at 1--5~Hz. The total 0.05--10~Hz variability has a
fractional rms amplitude of 20\%--45\%, well in excess of the 8\%--12\% rms broad-band noise usually seen in power spectra of SAX J1808.4--3658. 

We perform a detailed timing analysis of the flaring and study its relation to 
the 401~Hz pulsations. We find that the pulse amplitude varies proportionally with source flux
through all phases of the flaring, indicating that the flaring is likely due to
mass density variations created at or outside the magnetospheric boundary.
We suggest that this 1--5~Hz flaring is a high mass accretion rate version
of the 0.5--2~Hz flaring which is known to occur at low luminosity ($<13$~mCrab),
late in the tail of outbursts of SAX J1808.4--3658. 
We propose the dead-disk instability, previously suggested as the mechanism for the 
0.5--2~Hz flaring, as a likely mechanism for the high luminosity flaring reported here.
\end{abstract}

\keywords{
	pulsars: general -- 
	stars: neutron --
	X-rays: binaries --	
	individual (SAX J1808.4-3658)
}

\section{Introduction}
    The accreting millisecond X-ray pulsar (AMXP) SAX J1808.4--3658 (henceforth
    SAX J1808), was the first X-ray binary to show pulsations in the
    millisecond domain \citep{Wijnands1998}. Since its discovery with the
    BeppoSax satellite in 1996 \citep{Zand1998}, SAX J1808 has shown regular
    outbursts with a recurrence time of 2--3.5 yr. Between 1998 and
    2012 a total of 6 outbursts have been detected, all of which were
    extensively monitored with the {\it Rossi X-ray Timing Explorer} ({\it RXTE}). The
    light curves of these outbursts show a remarkably similar morphology
    \citep{Hartman2008}, consistently starting with a 2--5 day steep rise in
    flux (\textit{fast rise}), followed by a flattening that lasts for a few
    days (\textit{peak}). After reaching its peak X-ray luminosity, the light
    curve shows a slow exponential decay over a time span of several days
    (\textit{slow decay}), followed by faster linear decay
    (\textit{fast decay}) which typically lasts for 3--5 days
    \citep{Hartman2008}. After the fast decay, the source enters a prolonged, low
    luminosity state (\textit{outburst tail}), during which $\sim$5 day long episodes 
    of increased X-ray emission are seen \citep{Patruno2009}, causing the luminosity to
    vary between $5\E{32}$ and $5\E{35}$~erg~s$^{-1}$
    \citep{Wijnands2004, Campana2008}. After several weeks to months, the
    outburst tail ends and the source returns to a quiescent luminosity of
    $\sim5\E{31}$~erg~s$^{-1}$ \citep{Heinke2009}.
	
    Several types of variability are seen in SAX J1808. The 401~Hz coherent
    pulsations can be detected throughout the outburst, including the outburst
    tail \citep{Hartman2008, HartmanPatruno2009, Patruno2012}, and are thought
    to be caused by thermal emission from a localized region on the neutron
    star surface heated by the impact of plasma coming down the accretion
    funnel.  This so-called hotspot revolves with the neutron star spin,
    modulating the observed X-ray flux at the spin frequency. 
    The pulsations thus offer a physical tracer of the innermost accretion flow. 

    In addition to the pulsations, type I X-ray bursts \citep{Zand1998} and twin 
    kHz quasi-periodic oscillations \citep[QPOs;][]{Wijnands2003} have also been
    detected in SAX J1808. The stochastic variability of the 1998 and 2002
    outbursts of SAX J1808 has been studied by \citet{Straaten2005}, who find
    that the power spectral characteristics are similar to those of a typical
    atoll source. 
    
    A peculiar type of variability in SAX J1808 is the strong 0.5--2~Hz
    flaring often seen in the outburst tail at luminosities $<13$~mCrab.
    This phenomenon has frequently been designated as the `1~Hz QPO', however, in 
    this paper we refer to it as the `low luminosity flaring'. 
	The low luminosity flaring was first reported by \citet{Klis2000} and found to 
    occur sporadically throughout the outburst tails of 2000 and 2002 
    \citep{Wijnands2004}. An in-depth study by \citet{Patruno2009} showed that 
    it was also present in the tail of the 2005 outburst, but not in the tail 
    of the 2008 outburst. In the later 2011 outburst the low luminosity flaring
    was again not detected \citep{Patruno2012}, although due to Solar constraints 
	only the onset of the tail could be observed.
    
    The large luminosity variations in the outburst tails of SAX J1808 have been proposed to be caused by an
    intermittent propeller effect \citep{Campana2008}. In the propeller regime
    the inner edge of the accretion disk rotates slower than the neutron star
    magnetosphere and in-falling matter is no longer able to accrete onto the
    neutron star, but instead is blown out of the system as by a propeller
    \citep{Illarionov1975}. Considering the 0.5--2~Hz low luminosity flaring in the context of the 
    propeller onset, \citet{Patruno2009} propose the Spruit--Taam
    instability \citep{Spruit1993} as the most likely origin.  An open problem with
    this interpretation, however, is the sporadic nature of this flaring in the
    2000, 2002 and 2005 outburst tails as well as its complete absence in the 2008
    and 2011 outbursts. As noted by \citet{Patruno2009} other mechanisms such as an
    interchange instability \citep{Arons1976} can not be strictly ruled out.
	
    In the present work we report on the detection of strong 1--5~Hz flaring
    observed in the 2008 and 2011 outbursts of SAX J1808 that is similar to the
    0.5-2~Hz low luminosity flaring, but occurs at much higher luminosities, near 
    the peak of the outburst. We present the timing properties of this high luminosity
    flaring and discuss its nature. In Section~\ref{sec:DataReduction} we outline our data reduction
    procedure and our timing analysis methods. In Section~\ref{sec:Results} we
    present the results we obtained and in Section~\ref{sec:Discussion} we
    explore potential mechanisms for the high luminosity flaring and discuss our results
    in the context of the previously observed low luminosity flaring. Our conclusions are
    summarized in Section~\ref{sec:Conclusions}. 
	
\section{X-ray Observations} 
\label{sec:DataReduction}
    We used data collected with the {\it RXTE} Proportional Counter Array 
    \citep[PCA, see][for technical details]{Jahoda2006}. For the timing analysis we used 
    the pointed observations of all observed outbursts of SAX J1808, 
    excluding the outburst tail, selecting only data taken in GoodXenon or 
    122~$\mu$s Event mode. All data were binned to a time resolution of 1/8192~s 
    prior to further analysis.  
    
    Additionally, we used the 16 s time-resolution Standard-2 data to
    construct 2--16~keV X-ray light curves normalized to Crab and calculate the
    standard Crab-normalized soft (3.5--6.0~keV / 2.0--3.5~keV) and hard (9.7--16~keV
     / 6.0--9.7~keV) colors \citep[see e.g.][for the detailed
    procedure]{Straaten2003}. 
	
	\subsection{Stochastic Timing Analysis}	
	\label{sec:stochastic}
        To study the stochastic time variability we calculated Fourier
        transforms of 256 s data segments, giving a frequency resolution
        of $\sim4\E{-3}$~Hz and a Nyquist frequency of 4096~Hz. To
        optimize the signal-to-noise we selected only the events in channels 5--48
        ($\sim$2--20~keV). No background subtraction or dead-time correction was
        applied prior to the Fourier transform. We used the transforms to
        compute Leahy normalized power spectra and subtracted the Poisson level
        using the formula of \citet{Zhang1995} according to the method
        described by \citet{KleinWolt2004}. 

        Power spectra computed using segments of the same ObsID were averaged
        to improve statistics. Additionally, power spectra of consecutive
        ObsIDs were sometimes averaged to improve statistics further, but only
        if their flux and color properties were consistent with being the same
        and the power spectra were similar.		
        
        Finally, we normalized the averaged power spectra to source fractional rms 
        amplitude squared per Hz while correcting for the background count rate as 
        estimated using the \textsc{ftool} \textsc{pcabackest} \citep{Klis1995}. 
        In this normalization the fractional rms contribution, $r$, of a given 
        frequency band is 
		\begin{equation}
		\label{eq:rmssquared}
			r^2 = \int_{\nu_1}^{\nu_2} P(\nu)d\nu,
		\end{equation}
		where $P(\nu)$ is the power density in the described units.
        
        We describe the power spectra by fitting a function consisting of the
        sum of several Lorentzians plus a Schechter function for the flaring
        component (see Section~\ref{sec:Results}). We characterize each Lorentzian 
        by its centroid frequency and full-width-at-half-maximum (FWHM).
        The strength, or power, of the Lorentzian is obtained by integrating 
        it over frequency from 0 to $\infty$ and reported in terms
        of the corresponding fractional rms as defined in 
        equation~\ref{eq:rmssquared}.
        
        The Schechter function is an exponentially cut-off power law and
        is given by 
        $
			P(\nu) \propto \nu^{-\alpha}  e^{-\nu/\nu_{cut}}
		$
        \citep{Hasinger1989, Dotani1989}, with power-law index $\alpha$ and 
        cut-off frequency $\nu_{cut}$. We can define a 
		`centroid' frequency 
        $
        	\nu_0 = -\alpha\nu_{cut}
        $
        as the frequency at which the Schechter function reaches is maximum
        in $P(\nu)$.
        
	\subsection{Coherent Timing Analysis}
        We correct the photon arrival times to the solar system barycenter with the
        \textsc{ftool} \textsc{faxbary} using the source coordinates of \citet{Hartman2008}. 
        This tool also applies the {\it RXTE} fine
        clock corrections, allowing for high precision ($\sim$4 $\mu$s) timing
        analysis \citep{Rots2004}.  Selecting only the events in the channel
        range 5--37 ($\sim$2--16 keV) \citep{Patruno2012}, we fold the data to produce
        folded pulse profiles. 
        For the timing solution we use the ephemeris of \citet{HartmanPatruno2009} for
        the 2008 outburst and that of \citet{Patruno2012} for the 2011
        outburst. The length of the folded data segments varies depending on
        purpose; typically we fold $\sim$3000~s of continuous {\it RXTE} observations
        to obtain high signal-to-noise pulse profiles. Alternatively we take $\sim$256~s
        segments to obtain better time resolution when following the
        outburst evolution. In both cases we allow the segment length to vary
        such that we use the entire observed time series.

\section{Results} 
\label{sec:Results}
\begin{figure}
		\includegraphics[width=\colwidth]{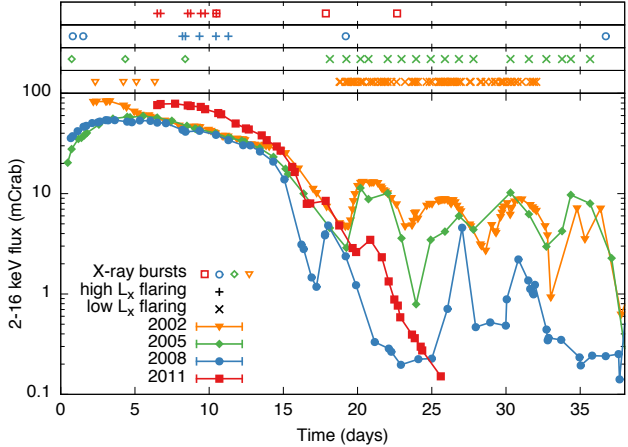}
        \caption{ 
        	Light curves in the 2--16~keV X-ray band of the 2002 (orange
			triangles), 2005 (green diamonds), 2008 (blue circles) and 2011 (red
			squares) outbursts. The symbols in the top bars show the times
			of type I X-ray bursts and observations with high luminosity flaring (pluses, this paper) 
			and low luminosity flaring \citep[crosses,][]{Patruno2009}. The light 
			curves have been shifted in time such that the transition from slow to 
			fast decay is at day 15, using a shift in MJD of 52553.6, 53522.3, 
			54731.2 and 55863.4 for the outbursts 2002 through 2011, respectively.}
		\label{fig:CombiLC}
	\end{figure}

    The outbursts of SAX J1808 show very similar light curves. In Figure~\ref{fig:CombiLC} 
    we show the four outbursts that occurred between 2002 and 2011. 
    Since the 1998 outburst does not show either flaring component and
    in 2000 only the  outburst tail could be observed, we do not show these outbursts and 
    exclude them from further analysis.
    
    In Figure~\ref{fig:CombiLC} the light curves where shifted in time such that the 
    transition from the slow decay to the fast decay occurs on day 15. The top panels 
    indicate the times of confirmed and suspected X-ray bursts (removed from the data prior to the analysis), 
    detections of the low luminosity flaring \citep{Patruno2009} and detections of 
    the high luminosity flaring (below).     

    Like the light curves, the power spectra of SAX J1808 are similar. 
    The power spectra can usually be described
    with 4 Lorentzians \citep{Straaten2005}; a broad noise `break' component at $\sim$4~Hz, a `hump'
    component in the range $\sim$20--80~Hz, a hectoHz component in the range
    $\sim$100--200~Hz, and the upper kHz QPO with a frequency in the range
    $\sim$300--700~Hz.
	
\begin{figure*}
		\includegraphics[width=\linewidth]{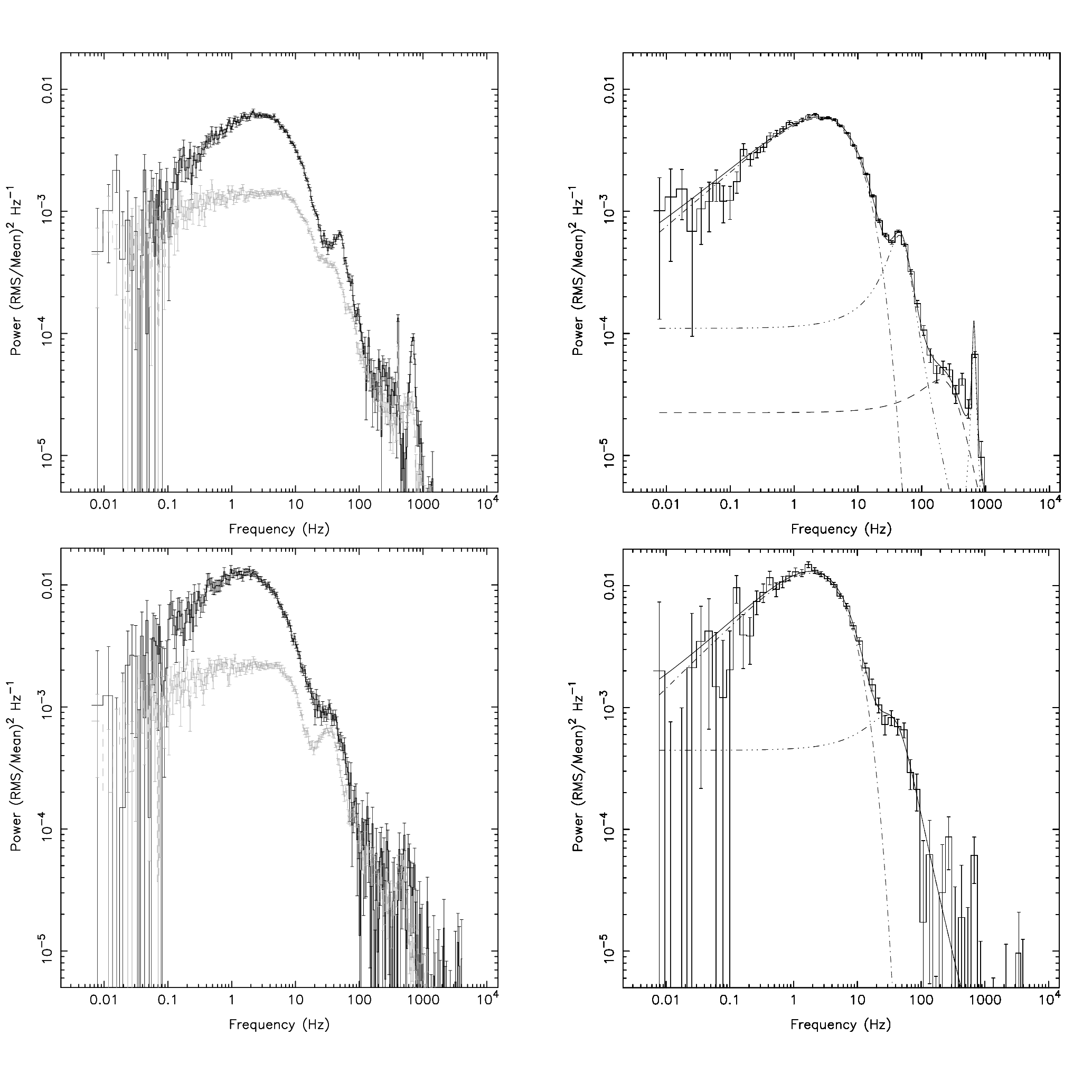}
        \caption{ 
        	Power spectra of the high luminosity flaring. The left column compares 
			the high luminosity flaring (average of all ObsIDs reported 
			Table~\ref{tab:ObsIds}) of 2011 (top, 
			black) and 2008 (bottom, black) with the broad band noise in the 2002 
			(top, gray) and 2005 (bottom, gray) outbursts in similar outburst stages 
			and spectral states. The right column shows the Schechter
            function fit to power spectra for the 2011 (top) and 2008 (bottom) outbursts.
            Note that all power spectra show power density versus frequency.}
		\label{fig:CombiPDS}
	\end{figure*}

    Both in 2008 and 2011 we find an interval spanning
    $\sim$3~days during which the power spectrum deviates significantly from
    its standard shape. The power density between 0.05~Hz and 10~Hz is much
    higher and rather than showing the usual horizontal plateau in $P(\nu)$,
    the power density shows a peak around $\sim$3~Hz (see Figure~\ref{fig:CombiPDS}). 
    Lorentzian and Gaussian functions fail to provide statistically acceptable
    fits to the power spectra as they are not steep enough at low frequencies. We 
    find that replacing the Lorentzian break component with a Schechter function does provide a
    satisfactory fit to the data.     
    The observations in which we see this
    behavior are marked with pluses in Figure~\ref{fig:CombiLC} and their ObsIDs are given in 
    Table~\ref{tab:ObsIds}.
    We combined the 12 observations showing this new phenomenon
    into 8 intervals and give all power spectrum fit parameters for those intervals 
    in Table~\ref{tab:QPOParameters}.
    
\begin{deluxetable}{lcccc}
	\tabletypesize{\scriptsize}
	\tablecaption{ 
		High Luminosity Flaring Observation Information
		\label{tab:ObsIds} 
	}
	\tablewidth{1.0\linewidth}
	\tablehead{
		\colhead{ObsID} & \colhead{Length (s)} & \colhead{Start MJD} & \colhead{Flux (mCrab)} & \colhead{rms (\%)$^a$}
	}
\startdata
\cutinhead{2008 Outburst}
A-07     & 3000 & 54739.40 & 43.2 & 28.5 \\
A-09     & 2300 & 54739.60 & 41.4 & 29.0 \\
A-02     & 6100 & 54740.54 & 42.2 & 27.0 \\
A-08     & 6100 & 54741.65 & 38.5 & 29.0 \\
B-00$^b$ & 1280 & 54742.49 & 34.1 & 38.7 \\
\cutinhead{2011 Outburst}
C-01 &  2800 & 55869.83 & 76.3  & 21.9 \\
C-00 & 12000 & 55870.16 & 78.2  & 21.7 \\
C-02 &  6100 & 55871.13 & 78.8  & 11.1 \\
C-03 &  5400 & 55871.98 & 75.7  & 22.4 \\
C-04 &  6100 & 55872.18 & 74.7  & 22.1 \\
C-05 &  3100 & 55872.86 & 73.2 & 18.9 \\
C-06 &  6100 & 55873.16 & 69.5  & 22.3 
\enddata
\tablenotetext{a}{0.05-10Hz fractional rms}
\tablenotetext{b}{Using only the first of three observation files in the ObsID.}
\tablecomments{A = ObsID 93027-01-02, B = ObsID 93027-01-03, C = ObsID 96027-01-01.}
\end{deluxetable}

\begin{deluxetable*}{lccccc}
	\tablecolumns{5}
	\tabletypesize{\scriptsize}
    \tablecaption{ 
    	Power Spectrum Fit Parameters
		\label{tab:QPOParameters} 
	}
	\tablewidth{\linewidth}
	\tablehead{
                & \colhead{high $L_x$ flaring} & \colhead{Hump}        & \colhead{hHz} & \colhead{upper kHz}
    }
\startdata
                & {$\alpha$}      & {FWHM}      & {FWHM}      & {FWHM}      \\
\colhead{Group} & {$\nu_{cut}$}   & {$\nu_{0}$} & {$\nu_{0}$} & {$\nu_{0}$} & {$\chi^2$/dof}\\
                & {$r$}           & {$r$}       & {$r$}       & {$r$}       \\
                        			
\cutinhead{2008}
            & $- 0.69 \pm 0.04 $  &  $ 45.6 \pm 4.9 $  &   \nodata   & $ 404  \pm 196 $ \\
A-07, A-09  & $  2.54 \pm 0.12 $  &  $ 21.4 \pm 2.8 $  &   \nodata   & $ 496  \pm  56 $ & 173/118 \\
            & $ 28.03 \pm 0.43 $  &  $ 29.1 \pm 1.8 $  &   \nodata   & $ 19.5 \pm 3.1 $ \\[0.15cm]

            & $- 0.41 \pm 0.02 $  &  $ 46.7 \pm 4.1 $  &   \nodata   & $ 147  \pm 103 $ \\
A-02        & $  4.21 \pm 0.15 $  &  $ 34.1 \pm 1.7 $  &   \nodata   & $ 618  \pm 31  $ & 148/121 \\
            & $ 28.72 \pm 0.29 $  &  $ 26.0 \pm 1.0 $  &   \nodata   & $ 12.1 \pm 2.3 $ \\[0.15cm]
            
            & $- 0.51 \pm 0.05 $  &  $ 60.1 \pm 9.7 $  &   \nodata   &   \nodata   \\
A-08        & $  3.59 \pm 0.24 $  &  $ 29.2 \pm 6.5 $  &   \nodata   &   \nodata   & 76.8/121  \\
            & $ 30.78 \pm 0.66 $  &  $ 28.6 \pm 2.7 $  &   \nodata   &   \nodata   \\[0.15cm]

            & $- 1.94 \pm 0.25 $  &  $ 14.1 \pm 3.2 $  &   \nodata   &   \nodata   \\
B-00        & $  1.40 \pm 0.19 $  &  $  9.1 \pm 1.6 $  &   \nodata   &   \nodata   & 102/75  \\
            & $ 34.4  \pm 1.6  $  &  $ 25.7 \pm 2.7 $  &   \nodata   &   \nodata   \\
\cutinhead{2011}
            & $- 0.45 \pm 0.01 $ & $ 40.8  \pm 3.1  $  &  $  427 \pm 232 $ & $  72   \pm 23   $ \\
C-01, C-00  & $  5.63 \pm 0.09 $ & $ 44.64 \pm 0.52 $  &  $  199 \pm 164 $ & $ 656.1 \pm  2.3 $ & 172/141 \\
            & $ 25.48 \pm 0.10 $ & $ 20.20 \pm 0.92 $  &$16.7_{-2.3}^{+11.2}$&$11.54 \pm 0.76 $ \\[0.15cm]

            & $- 0.79 \pm 0.15 $ & $ 58   \pm 18  $  &   \nodata   &  $ 494  \pm 430 $ & \\
C-02        & $  5.65 \pm 0.95 $ & $ 24   \pm 13  $  &   \nodata   &  $ 466  \pm 118 $ & 58.9/65 \\
            & $ 15.4  \pm 1.3  $ & $ 13.7 \pm 3.3 $  &   \nodata   &  $ 12.9 \pm 3.8 $ & \\[0.15cm]

            & $- 0.55 \pm 0.02 $ & $ 52.0  \pm 3.3  $  &  $ 151  \pm 50   $  &  $  95    \pm 17    $  \\ 
C-03, C-04  & $  5.29 \pm 0.11 $ & $ 46.4  \pm 1.1  $  &  $ 321  \pm 24   $  &  $ 716.0  \pm  4.6  $ & 228/123  \\
            & $ 26.75 \pm 0.15 $ & $ 21.48 \pm 0.54 $  &  $  8.8 \pm  1.2 $  &  $  13.34 \pm  0.64 $ \\[0.15cm]

            & $- 0.59 \pm 0.02 $ & $ 31.3  \pm 4.6  $  &  $ 472  \pm 255 $  &  $  61   \pm 37    $ \\
C-05, C-06  & $  5.80 \pm 0.13 $ & $ 50.04 \pm 0.82 $  &      0 (fixed)     &  $ 709.3 \pm  3.3  $ & 140/124 \\ 
            & $ 27.31 \pm 0.18 $ & $ 16.0  \pm 1.4  $  &  $ 21.6 \pm 1.9 $  &  $ 12.60 \pm  0.74 $ 
\enddata
\tablecomments{A = ObsID 93027-01-02, B = ObsID 93027-01-03, C = ObsID 96027-01-01. Frequencies $\nu_0$ and 
$\nu_{cut}$ and FWHM are in units of Hz, and fractional rms, r, is expressed as a percentage.}
\end{deluxetable*}

\begin{figure}
		\includegraphics[width=\colwidth]{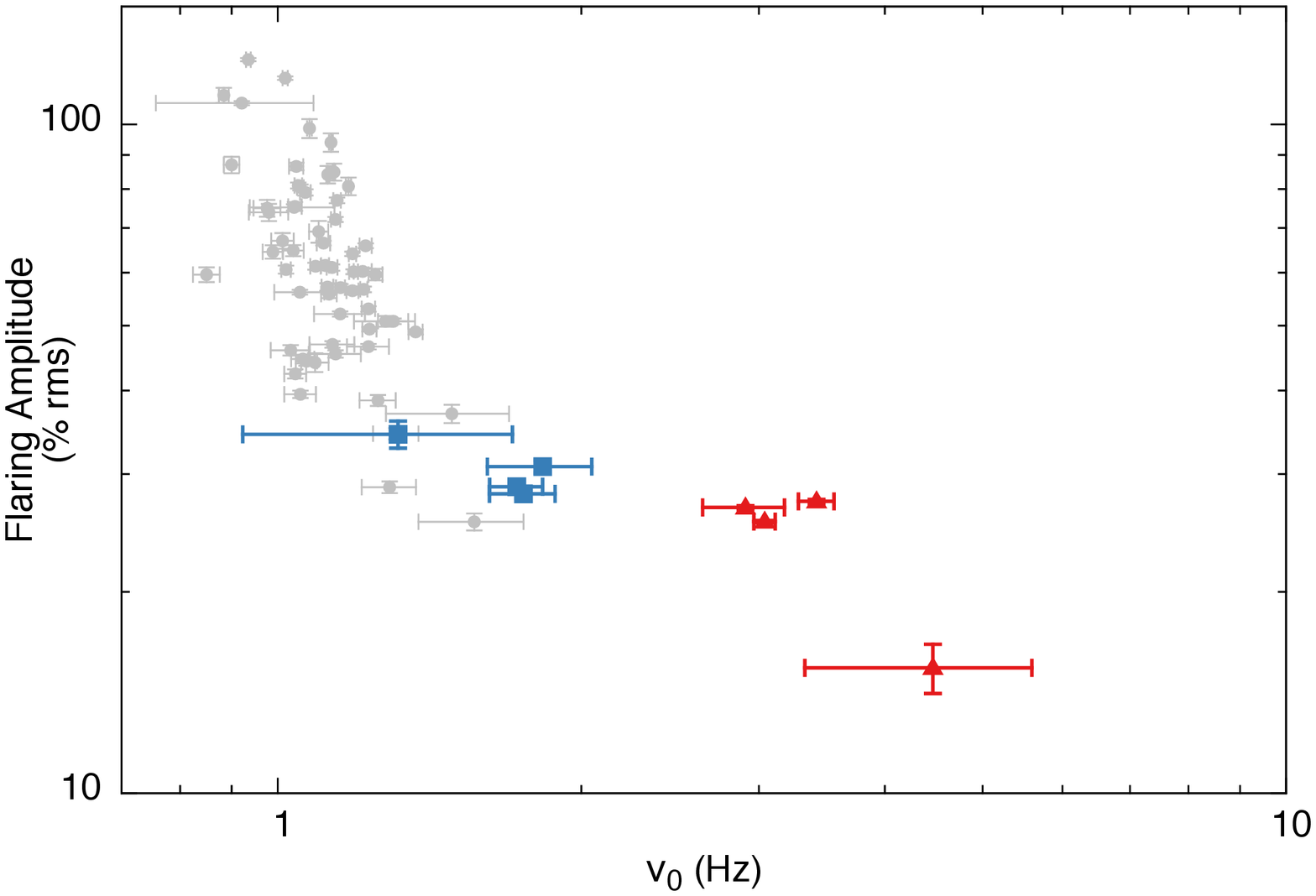}
        \caption{
        	Correlation between the fractional rms and the high and low
			luminosity flaring centroid frequencies. We show the 2008 (blue squares) 
			and 2011 (red triangles) high luminosity flaring frequency obtained in this work
			along with the low luminosity flaring frequencies (grey) from \citep{Patruno2009}
			against their respective fractional rms.
		}
		\label{fig:QPOrelations}
	\end{figure}

	The centroid frequency of the best-fit Schechter function, as defined in
	Section~\ref{sec:stochastic}, varies between 1 and 5~Hz.
    The amplitude of the flaring varies between 26\% and 34\% rms, and shows a clear 
    anti-correlation with $\nu_0$ (see Figure \ref{fig:QPOrelations}).
    The power-law index $\alpha$ shows some scatter, but no correlation
    with frequency. 
    
\begin{figure*}
		\includegraphics[width=1\linewidth]{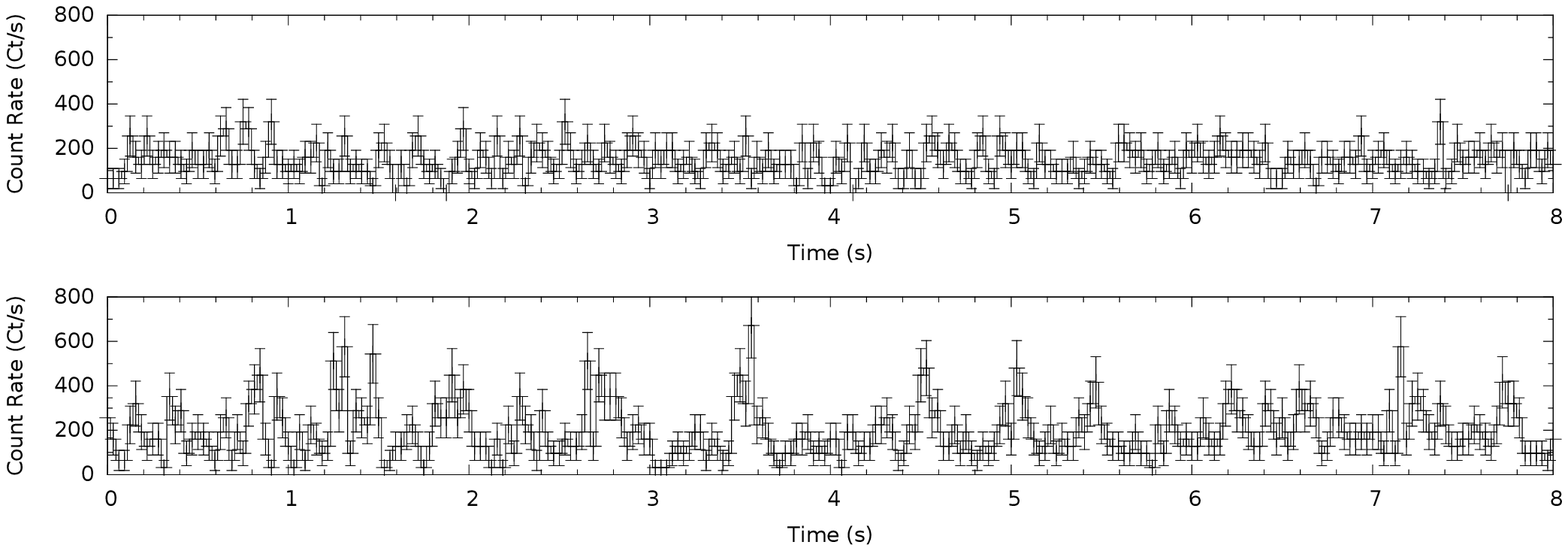}
        \caption{
	        Light curve segments of the 2008 outburst at a time resolution of 0.035~s. The 
	        top panel shows a segment of ObsID 93027-01-03-02 which shows the regular broad band noise in
	        the power spectrum. The bottom panel shows a segment of ObsID 93027-01-03-00,
	        during the time that the high luminosity flaring shows an amplitude of 45\% rms.
	    }
        \label{fig:FlaringLC}
	\end{figure*}

    The morphology of this new component is very different from the regular
    broad band noise, but quite similar to that of the low luminosity flaring
    seen in the outburst tails. Because the power spectrum returns to its normal 
    shape outside the reported $\sim$3~day intervals, we suggest that this new
    high luminosity flaring phenomenon dominates over the
    Lorentzian break component normally seen in the same frequency range.
    During the 2008 outburst the amplitude of this new component briefly reaches $\sim$45\% rms
    (see Section~\ref{sec:2008}). In the bottom panel of Figure~\ref{fig:FlaringLC} we show 
    a small selection of the light curve during this time. The $\sim$3~Hz 
    variability is directly visible and indeed has a flaring morphology, showing 
    relatively sharp spikes of emission at quasi-regular separation. 
    Contrary to the 0.5--2~Hz \textit{low luminosity flaring}, which is seen only
    in the outburst tail for luminosities $<13$~mCrab, the 1--5~Hz flaring, reported here
    for the first time, occurs near the peak of the outburst at luminosities $>30$~mCrab and is therefore
    called the \textit{high luminosity flaring}.
    	
    In order to study the evolution of the high luminosity flaring on short time scales
    we compute power spectra of 256~s data segments. Fitting a Schechter
    function to these short segment power spectra does not provide meaningful
    results, so instead we characterize the high luminosity flaring power by integrating
    the power spectrum between 0.05 and 10~Hz, the range in which we observe 
    the bulk power in excess of the expected break component. 
    These frequency bounds are identical to those used by
    \citet{Patruno2009}, allowing a comparison of the low luminosity flaring 
    results with the results we obtain for the high luminosity flaring. The power 
    estimates obtained in this way are described below. They match with 
    the power measured by fitting a Schechter function to
    power spectra of longer observations. 

    We also study the relation of the high luminosity flaring with the 401~Hz
    pulsations by constructing pulse profiles for the same data segments
    and considering the joint behavior of the flaring rms amplitude and the
    phase and amplitude of the pulsations. We now discuss these results in
    detail for each outburst.

	\subsection{2008}
	\label{sec:2008}
\begin{figure*}
			\includegraphics[width=\linewidth]{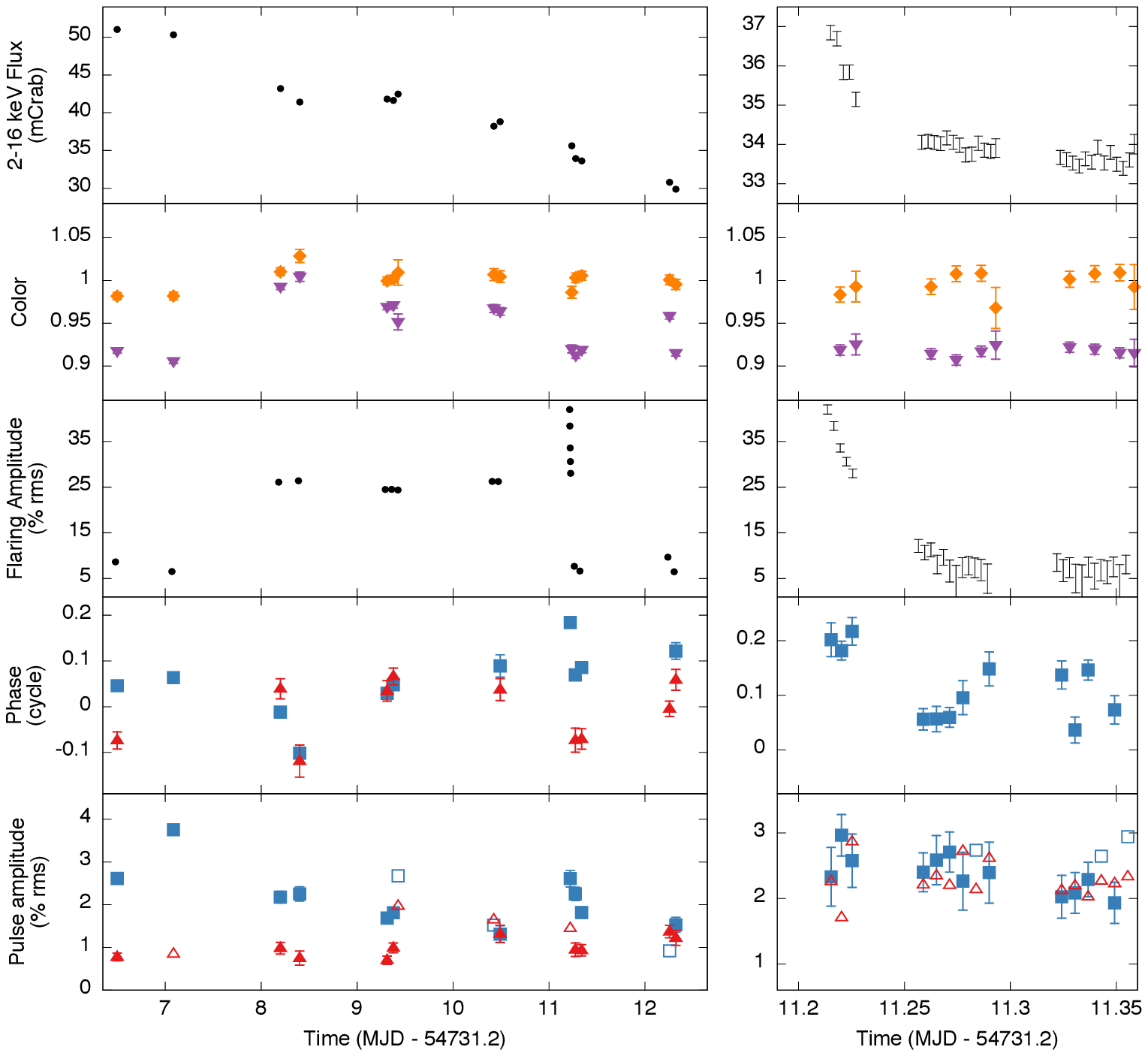}
	        \caption{
				Evolution of the high luminosity flaring in the 2008 outburst,
				showing in the left column: the 2--16 keV light curve (top panel); 
				soft (purple triangle) and hard
                (orange diamond) colors (second panel); 0.05--10~Hz high
                luminosity flaring fractional rms (middle panel); the pulse
                phases (fourth panel) and finally the pulsed fractional rms
                (bottom panel). Amplitude and phase measurements of the
                pulsations are given for the fundamental (blue squares) and the
                second harmonic (red triangles). Open symbols give 95\%
                confidence upper limits. The right column shows the same data,
                but zoomed in on the flaring switch-off.
                }
			\label{fig:Panels2008}
		\end{figure*}

        In 2008 the high luminosity flaring is seen at the onset of the slow decay. The 
        flaring is present for $\sim$3.1 days between MJD 54739
        and MJD 54742, during which the flux decays from 43 to 34 mCrab. 
        In Figure \ref{fig:Panels2008} we present the
        source evolution during this interval showing, from top down, the 2--16~keV 
        light curve, hard (orange) and soft (purple) colors, 0.05--10~Hz
        rms amplitude, and the phase and amplitude for the
        fundamental (blue) and second harmonic (red) of the 401~Hz
        pulsations. In the left column we averaged the data segments per
        $\sim3000$~s {\it RXTE} observations and the right column uses the full
        256~s resolution.
        We describe the evolution as a function of days using the time shift
        of Figure~\ref{fig:CombiLC}. 
        
        The flaring is first observed at day 8.3 with a fractional rms of 28\%.
        The previous observation $\sim$1.1 days earlier, showed a fractional
        rms of 8\%. Assuming this level of fractional variability due to the
        break component persists incoherently during the flaring interval, an
        rms amplitude of 27\% is deduced for the flaring. 
         
        Between day 8.3 and 10.3 the flaring amplitude varies between 25\% and 28\% rms, 
        showing a weak anti correlation with flux on a day-to-day
        timescale.  In the same period the soft color slowly decreases while
        the hard color remains roughly constant.  Since these color trends
        extend beyond the flaring interval, and are also seen in the same stage
        in the other outbursts \citep{Straaten2005}, they are most likely not
        related to the presence of flaring.
        During the flaring interval the pulse amplitude and phase show
        complicated variations, again also seen outside the flaring interval
        and in the other outbursts \citep[see e.g.][]{Hartman2008}, which
        relate to the flux variations rather than the presence of flaring
        \citep{Patruno2009b}.

        At day 11.3 we see the flaring at an amplitude of 45\% before it
        switches off. During the switch-off we see the flaring rms drop
        dramatically over the course of 5 consecutive data segments, showing a
        strong correlation with a small drop flux. 
        A zoom-in of this entire episode is shown in the right column of in
        Figure~\ref{fig:Panels2008}.  Assuming the flaring started to switch
        off at the highest observed fractional rms and completely turned off at
        day 11.27, we obtain a switch-off time of 1.2 hr. Since the decay
        could have set in prior to the observed peak this is a lower limit.
		        
        While the flaring rms amplitude drops rapidly, the pulse fundamental
        remains steady in amplitude and phase. Yet, at day 11.27, just as the
        fractional rms returns to the 8\% base level, there is a systematic
        drift in the fundamental phase of about 0.1 cycles, which is does not
        relate to a flux variation and might be a response to the flaring
        mechanism.

	\subsection{2011}
\begin{figure*}
			\includegraphics[width=\linewidth]{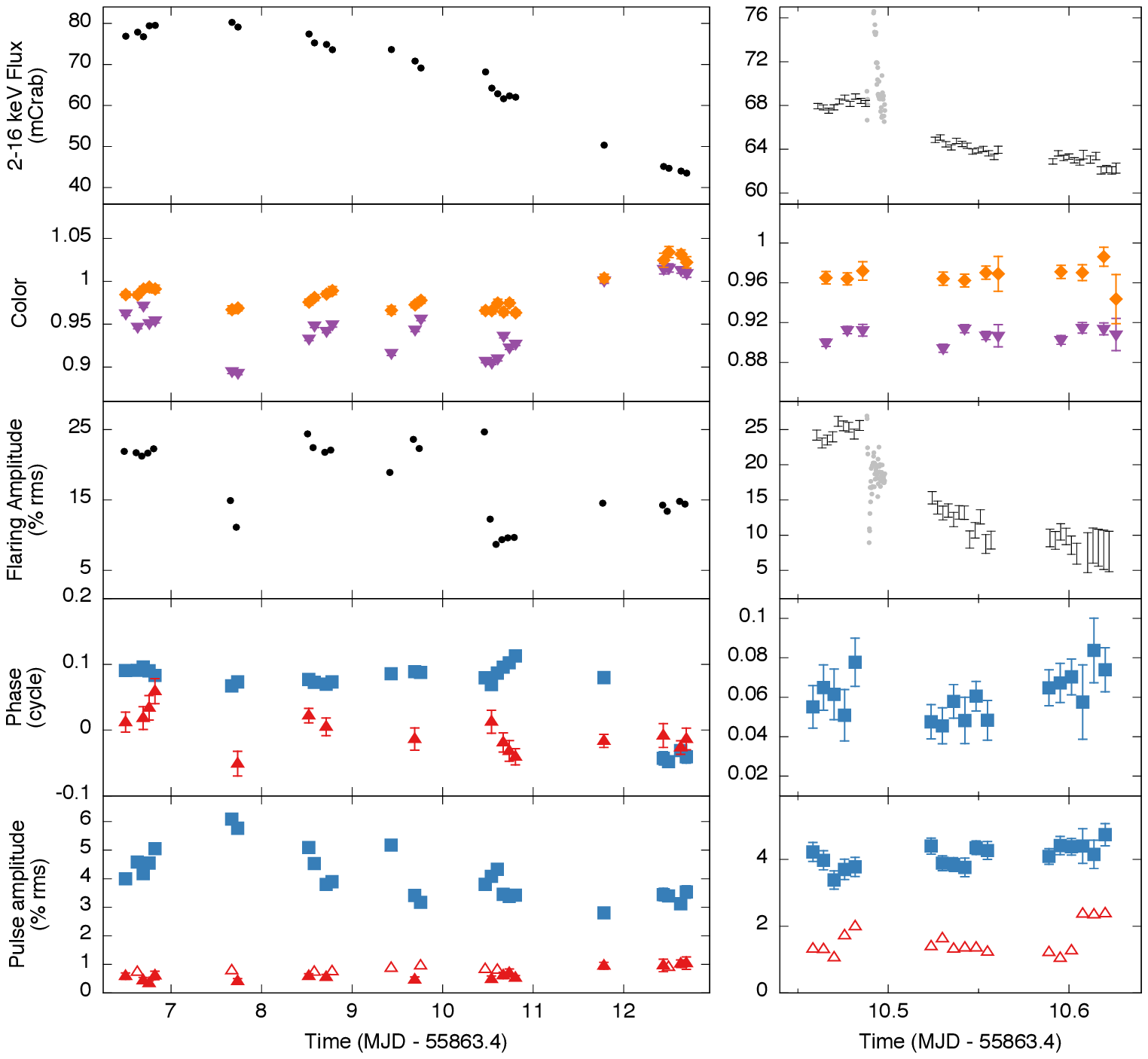}
	        \caption{ 
                Evolution of the high luminosity flaring in the 2011
                outburst in comparison to the 401~Hz pulsations. The grey bullets
                in the right column show the flux and 0.05--10~Hz fractional
                rms of 16~second data segments during a type I X-ray burst interval.
                For a detailed view see Figure~\ref{fig:BurstPanels}. For details
                on the remaining data see Figure~\ref{fig:Panels2008}.
			}
			\label{fig:Panels2011}
		\end{figure*}

        In Figure \ref{fig:Panels2011} we show the outburst evolution during
        the flaring interval of the 2011 outburst.  In the first observation of
        the 2011 outburst, at MJD 55869.9, the flaring is already present with
        a 0.05--10~Hz rms of $\sim23\%$.  The flaring stays active for the
        following 4~days, during which the flux decays from $76$ to $69$~mCrab.
		
        At day 7.9 the flaring decreases in strength, with the amplitude
        dropping consecutively to $\sim15\%$ and $\sim9\%$ rms.  The rms drop
        coincides with the outburst peak luminosity, and an unexpected drop in
        soft color. At the same time the pulsations show a drop in the second
        harmonic phase and an increase in the fundamental amplitude, however,
        this could again be related to the flux rather than the flaring.
        
        Over the entire flaring interval, between days 6.5 and 10.5, the
        flaring rms varies between 17\% and 26\%, showing a correlation with flux
        on a 3000~s timescale and an anti-correlation on the longer day-to-day
        timescale. 
        
\begin{figure}
			\includegraphics[width=\colwidth]{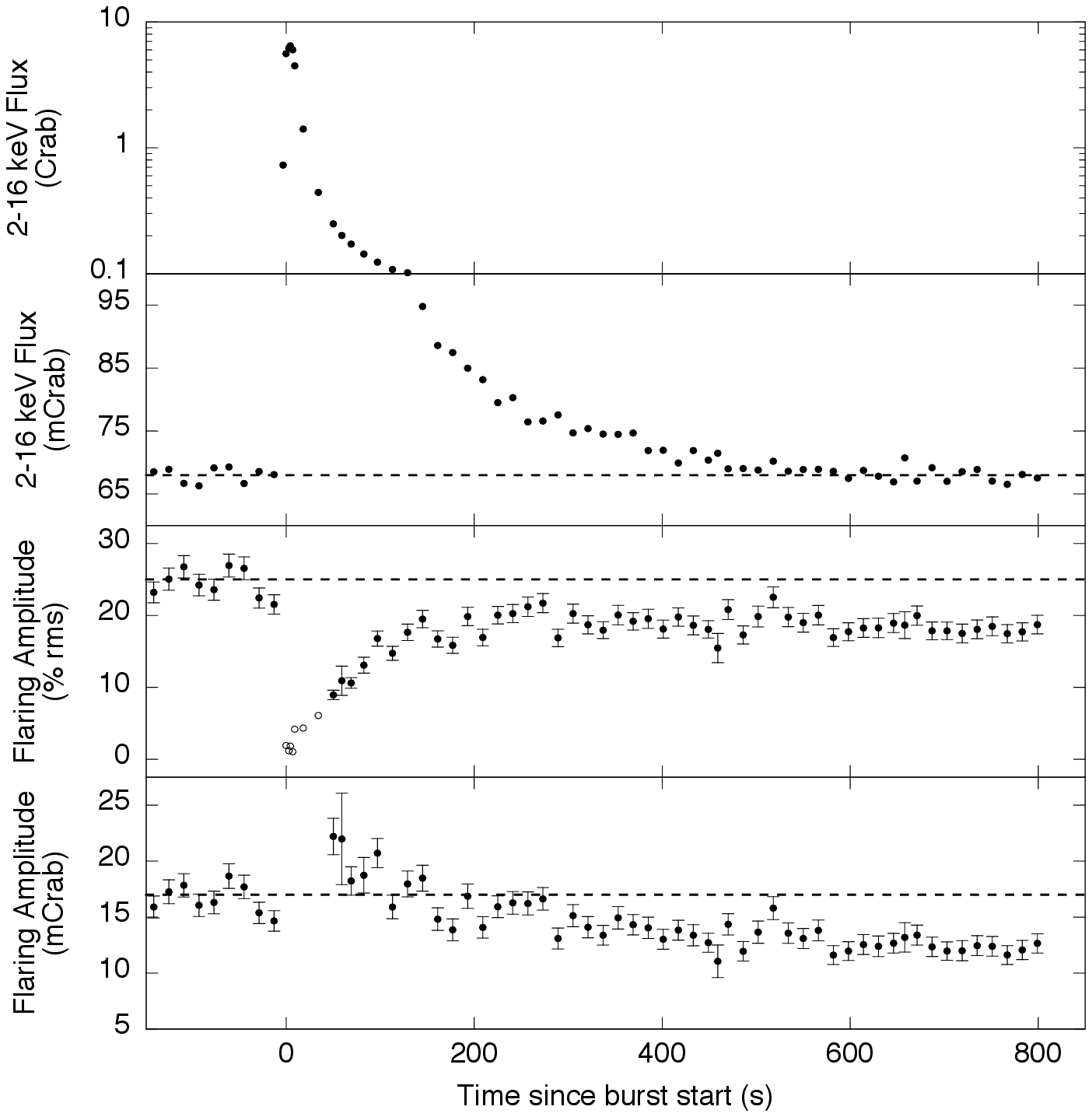}
	        \caption{ 
                Evolution of the 2--16~keV X-ray flux (top, second), the
                0.05--10~Hz fractional rms (third) and absolute rms (bottom)
                of the 2011 type I X-ray burst in seconds since the start of the
                burst. The open symbols in the third panel give 95\% C.L. upper
                limits on the fractional rms. The corresponding upper limits
                on the absolute rms where greater than 50~mCrab and have been omitted
                for clarity. Horizontal dashed lines show the respective averaged
                pre-burst levels. 
			}
			\label{fig:BurstPanels}
		\end{figure}

        At day 10.49 a bright type I X-ray burst occurs \citep{Zand2013}.
        Prior to the X-ray burst the flux is constant and the flaring amplitude
        holds steady at 25\%~rms (see Figures~\ref{fig:Panels2011} and~\ref{fig:BurstPanels}). 
        During the X-ray burst peak flux ($\sim7$~Crab) there is no significant
        variability in the 0.05--10~Hz frequency band above the Poisson noise level, giving 
        a 95\% confidence level upper limit on the flaring amplitude of 2\%~rms. As the burst 
        flux decays, the 0.05--10~Hz fractional rms slowly increases, such that the
        absolute rms is consistent with the pre-burst level. At 550~s after 
        the X-ray burst went off, just before the end of the {\it RXTE} observation, the 
        flux returns to its pre-burst level, but at 18\%~rms the flaring amplitude 
        does not. 
        
        In the next {\it RXTE} observation, at day 10.52, the flaring amplitude is
        15\% rms, and shows a steady decrease until it reaches the 8\% rms of 
        the broad band noise on day 10.6 . Outside the X-ray burst interval, 
        the switch-off is again strongly correlated with a
        small drop in flux.  Assuming the switch-off started when the X-ray
        burst occurred, we obtain a switch-off time of $\sim2$~hr. Like in
        the 2008 outburst, the pulse fundamental phase shows a systematic drift
        of 0.1 cycles after the flaring switched off, although for the 2011
        outburst this drift is much slower.  
		
	\subsection{ Summary of Observed Correlations }
	\label{sec:correlations}
        The sharp decrease in flaring amplitude when the flaring switches off
        in both outbursts and the rms drop at peak flux in 2011 suggest the
        flaring only occurs in a specific flux range with a sharp upper and
        lower bound which differs between the two outbursts. The flaring
        switch-off takes place on a timescale of 1--2~hr and is strongly
        correlated with flux. In one outburst the start of the flaring
        switch-off coincides with a bright type I X-ray burst. 
		
        Directly after the flaring switches off and the 0.05--10~Hz rms returns
        to the 8\% rms of the broad band noise, the pulse phase of the
        fundamental shows a systematic drift of 0.1 cycles for both outbursts. 

        For both outbursts the flaring fractional amplitude correlates with
        flux on a 3000~s timescale, while on a longer day-to-day timescale the
        flaring fractional amplitude is anti-correlated with flux. If we
        consider the relation of the 0.05--10~Hz fractional rms versus mean flux
        in the 2002 and 2005 outbursts, we find that the broad band noise
        follows similar trends, but with different correlation factors, indicating
        the observed relations between flux and flaring rms are not due the underlying
        broad band noise. This supports our assumption that the flaring is a distinct 
        component that is added onto the broad band noise.

	\subsection{Relation of the Flaring Phase with Pulsations}
	\label{sec:FlarePhase}
		In ObsID 93027-01-03-00 (day 11.2 of the 2008 outburst) the flaring 
		reached an rms of 45\% and was directly visible in the light curve 
		(Figure~\ref{fig:FlaringLC}). This allows us to investigate the pulsations 
		as a function of the flux variations at the flaring timescale.

		We construct a light curve at 1/32~s time resolution and sort the light
		curve in flux quartiles. To account for longer timescale variations in
		flux we apply the quartile selection on short 8~s segments containing
		256 flux estimates. We then construct 4 pulse profiles corresponding to the
		the 4 flux quartiles. We find that the pulse phases in different quartiles are 
		statistically consistent. The absolute amplitude of the pulsations is found
		to be proportional to flux ($\chi^2/\mbox{dof}=0.6$), such that the fractional 
		amplitudes are the same within the errors (see Table~\ref{tab:PulseFluxVariations}).
		
\begin{deluxetable}{ lccccc}
	\tabletypesize{\scriptsize}
	\tablecaption{ 
		Pulse Parameters for Quartile Selection
		\label{tab:PulseFluxVariations} }
	\tablewidth{1.0\linewidth}
	\tablehead{
		\colhead{Count rate} & \colhead{Pulse Amplitude (\% rms)} & \colhead{Error} & \colhead{Phase} & \colhead{Error}
	}
\startdata
    70.6   &  2.55  &  0.83  &  0.27  &  0.05  \\
    138.3  &  2.09  &  0.57  &  0.30  &  0.04  \\
    193.1  &  2.53  &  0.39  &  0.26  &  0.02  \\
    314.8  &  2.93  &  0.33  &  0.30  &  0.02 
\enddata
\end{deluxetable}

	\subsection{Energy Dependence}
	\label{sec:rmsSpectrum}
\begin{figure}
			\includegraphics[width=\colwidth]{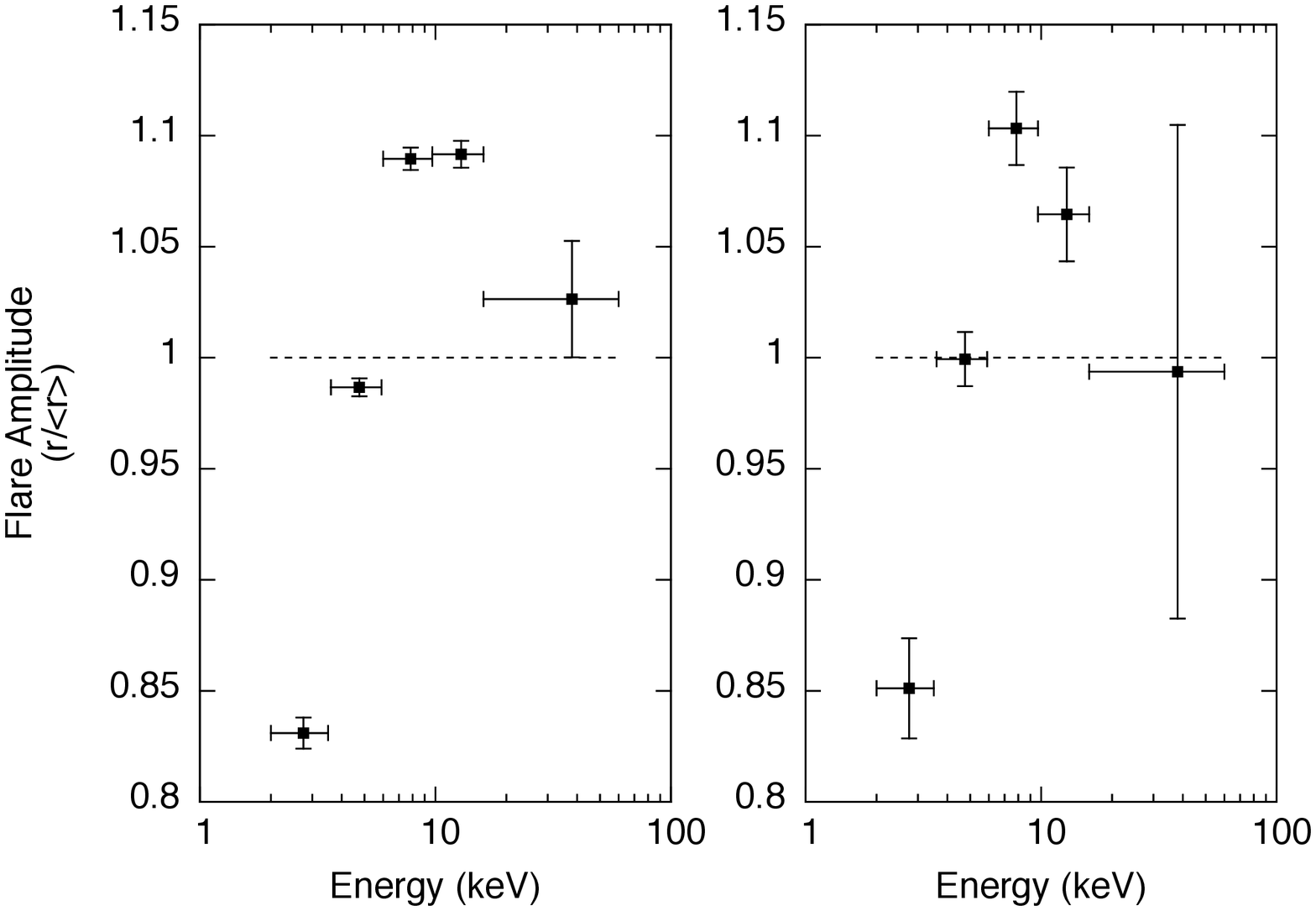}
	        \caption{
	        	Fractional rms spectra for flaring (right) and non-flaring (left) intervals
				in the 2011 outburst. We show the fractional rms divided by the mean fractional
				rms of the full energy band (22.1\% and 13.3\%, respectively).	        
	        }
			\label{fig:RmsSpectrum}
		\end{figure}

		Figure~\ref{fig:RmsSpectrum} shows the energy dependence of the high luminosity flaring
		fractional rms as measured in the 0.05--10~Hz band. We consider the fractional rms,
		which means the rms spectrum is divided by the energy spectrum of the mean flux. If the 
		flaring has the same energy dependence as the mean flux, then we
		expect the fractional rms to be constant in energy at the level of the fractional 
		rms of the full energy band. We find that the rms energy spectrum deviates from
		the average spectrum by contributing more between 6--16~keV and less below 6~keV.
		
		This harder-than-average spectrum suggests that the high luminosity flaring
		does not originate from the soft disk emission, but instead comes from a harder spectral
		component, possibly the neutron star surface or boundary layer
		\citep{Gilfanov2003}.
		
		We observe the same fractional rms spectral shape for the flaring during both 
		the 2008 and 2011 outbursts. More importantly, in this frequency range the same
		rms spectral shape is also found for observations where the flaring is not seen, which suggests that the
		broad band noise in the 0.05--10~Hz range and the flaring share a similar origin. 
		Since the broad band noise is believed to be caused by mass accretion rate variations in the inner
		accretion disk, this suggests that the high luminosity flaring is an accretion rate variation
		as well. 
		
\section{Discussion} 
\label{sec:Discussion}
    We have found two intervals of unusual 1--5 Hz flaring occurring at high luminosity 
    ($>30$~mCrab) in the 2008 and 2011 outbursts of SAX J1808 as observed with {\it RXTE}.
    This high luminosity flaring is characterized by spikes of emission in the light curve 
    at quasi-regular intervals, which result in a broad noise component in the 
    power spectrum. It is only seen for 3--4 days per outburst in a narrow $\sim$10~mCrab wide
    flux window at a different flux for 2008 and 2011, and only in those 2 (of 6) outbursts.
    
    The characteristics of the high luminosity flaring are reminiscent of
    those of the low luminosity ($<13$~mCrab) flaring previously detected in the tail of the
    of 2000, 2002 and 2005 outbursts \citep{Patruno2009}. The low luminosity flaring also
    produces a broad noise component in the power spectrum and is also seen exclusively in 
    a narrow $\sim10$ mCrab wide flux window, which, however, is at the same flux for all three outbursts. 
    The high and low luminosity flaring also differ in a number of other ways. 
    Specifically, the high luminosity flaring is seen at $\sim$35 and $\sim$75~mCrab, 
    with centroid frequency, $\nu_0$, of 1--5~Hz and rms amplitudes of 20--45\%. The low luminosity 
    flaring, on the other hand, is seen between 2 and 13 mCrab, with $\nu_0$ of 0.5--2~Hz and rms amplitudes 
    between 40\% and 120\%. As seen in Figure~\ref{fig:QPOrelations} the trend in
    fractional rms against frequency of the high and low luminosity flaring match at $\sim$40\% rms,
    but the relation is different for the two types of flaring. Additionally, the absolute
    rms increases with frequency. 
    
    The similarities between the high and low luminosity flaring suggest they could
    be caused by the same mechanism, however, the large discrepancy
    in the luminosity at which they are detected might indicate the opposite. We
    therefore first consider mechanisms that can explain the properties we observe
    for the high luminosity flaring alone, then we consider models that could explain them
    commonly with the low luminosity flaring.
    
    \subsection{Flaring Origin}
    	The observed flaring can either be caused by an extra emission component from 
		the accretion disk or neutron star; or by periodic obscuration along the line
		of sight or a variation in the accretion flow.
		
		\subsubsection{An Isolated Component}
		\label{sec:rmslimit}
		In both the 2008 and 2011 outbursts the flaring switched off smoothly, showing a strong
		correlation between a drop in flux and the decreasing flaring amplitude.
		This correlation might suggest that the flaring is due to some isolated emission
		component that is added to the mean flux.
		For such a process, we can estimate
		an upper limit on fractional rms of the total signal by assuming the extra component
		is 100\% modulated. We consider a signal of the form $f(t) = a + p(t)$, such
		that $a$ gives the unperturbed flux and $p(t)$ the added flaring signal. The average
		of $f(t)$ is $\mu_f = a + \mu_p$ with $\mu_p$ the average of $p(t)$. 
		The fractional rms of the total signal is defined as
		\begin{equation}
			r = \frac{\mbox{rms}}{\mu_f} = \frac{ \sqrt{ \int_0^T (f(t) - \mu_f)^2 dt / T } }{\mu_f},
		\end{equation}
		with $T$ the average period of a single flare. 
		For simplicity we take p(t) to be a square wave with duty cycle $D$, such that 
		wave amplitude is $\mu_p/D$. The enumerator now 
		reduces drastically, giving
		\begin{equation}
		\label{eq:dutycyclerms}
			r = \sqrt{D^{-1}-1} \ \frac{ \mu_p }{ \mu_f }.
		\end{equation}
		For $\mu_f$ we can use the observed flux prior to switch-off, which is
		37 and 68~mCrab for 2008 and 2011 respectively. The flux drop during the
		switch-off gives $\mu_p$, which is 3~mCrab in both outbursts. To calculate 
		the minimum required duty cycle we use the maximum observed fractional rms 
		(45\% and 26\%, respectively) and find that a duty cycle of $\sim0.03$ is needed. 
		This is much higher than the $D\simeq0.25$ that is apparent in the light curve
		(see e.g. Figure~\ref{fig:FlaringLC}), so the flaring cannot be produced 
		by an isolated emission process.
		
		\subsubsection{ Surface Processes }		
    	In Section~\ref{sec:FlarePhase} we found that the pulse amplitude responds
		to flux variations due to individual flares. This might suggest that the flaring 
		originates from the neutron star surface, for instance from QPOs 
		on the nuclear burning timescale in the hotspot \citep[see e.g.][]{Bildsten1998,Heger2007}.
		
		Marginally stable nuclear burning occurs in the transition between stable and unstable
		burning. It therefore occurs only in a narrow range of mass accretion rates, naturally
		explaining the narrow flux window observed for the high luminosity flaring. Furthermore,
		a transition to unstable burning, a type I X-ray burst, should stop the flaring,
		which is indeed observed during the 2011 flaring switch-off.
		
		The predicted oscillation timescale of marginally stable burning depends
		on the thermal and accretion timescale of the burning region and is $\sim100$~s
		\citep{Heger2007},
		which is too slow to account for the observed flaring. Furthermore,
		the flux variation due to the oscillations is limited by the fraction of
		nuclear energy and gravitational energy released per accreted nucleon. This fraction is of 
		the order of a few percent, resulting in a fractional rms of $\sim5-10\%$ (depending
		on duty cycle, see equation \ref{eq:dutycyclerms}), far too low to account for the large fractional amplitude of
		the high luminosity flaring. Nuclear burning therefore is an unlikely candidate.

		\subsubsection{Obscuration}
		Another possibility is that the flaring is caused by periodic obscuration
		of the neutron star. Dipping low-mass X-ray binaries show QPOs in the 
		$\sim$0.5--2.5~Hz range at rms amplitudes up to 12\% \citep{Homan1999, Jonker1999, 
		Jonker2000}. These QPOs were proposed to be caused by a nearly opaque medium
		orbiting at the radius where the orbital frequency matches the observed QPO
		frequency. This argument is supported by the flat rms spectrum of the QPO,
		the high inclination angle of the binaries, and constant fractional rms amplitude
		during short term luminosity variations like dips and type I X-ray bursts, which 
		are all characteristics of obscuration.
		
		The frequency range of the dipping QPO is similar to that of the high luminosity
		flaring, but all other characteristics differ. SAX J1808 does not show dips in
		the light curve and at an inclination of $60^{\circ}$ \citep{Cackett2009,
		Ibragimov2009,Kajava2011} is unlikely to show obscuration. The rms spectrum
		of the flaring is not flat, and the flaring fractional rms is diluted by
		the increased flux on a type I X-ray burst. An obscuration origin for the 
		high luminosity flaring can therefore be ruled out.

		\subsubsection{ Accretion Flow Variations }
		The rms spectrum (Section~\ref{sec:rmsSpectrum}) suggests the high luminosity flaring 
		may be due to variations in the accretion flow. If this is the case, flaring variability 
		is present in the accretion flow channeled to the hotspot, and should therefore
		affect both the persistent emission and the pulsed emission in a similar fashion.
		This relation between the flares and the pulsations was indeed observed (see
		Section~\ref{sec:FlarePhase}).		
		As, therefore, the accretion flow is the most plausible origin for the high
		luminosity flaring, we consider a number of such mechanisms in greater detail.
            
    \subsection{Disk Length- and Timescales}
    	We discuss instabilities in the accretion flow within the framework of the interplay between
		the corotation and magnetospheric radius. The corotation radius, $r_c$, is defined
		as the radius where the disk Keplerian frequency matches the neutron star spin 
		frequency and the magnetospheric radius, $r_m$, as the radius where magnetic stresses 
		equal the material stresses of the accreting plasma. 
		For specificity we note that $r_c$ can be written as
		\begin{equation}
			r_{c} \simeq 31 
					\left[\frac{\nu_{s}}{400 \mbox{ Hz}}\right]^{-2/3}
					\left[\frac{M}{1.4 \Msol}\right]^{1/3} \mbox{ km},
		\end{equation}
		with $\nu_s$ the spin frequency and $M$ the neutron star mass. Additionally we write $r_m$ 
		as \citep{Spruit1993,dAngelo2010}
		\begin{align}
			r_{m} \simeq 20 
					&\left[ \frac{M}{1.4\Msol} \right]^{-1/10}
					\left[ \frac{B}{10^8 \mbox{ Gauss}} \right]^{2/5}
					\left[ \frac{R}{10 \mbox{ km}} \right]^{6/5}
					\\ \nonumber
					&\times
					\left[ \frac{\dot{M}}{4\E{-10} \Msol\mbox{yr}^{-1}} \right]^{-1/5}
					\left[ \frac{\nu_s}{400 \mbox{ Hz}} \right]^{-3/10}
					\mbox{ km},
		\end{align}
		where $B$ is the magnetic field strength, $R$ the neutron star radius and
		$\dot{M}$ the mass accretion rate.
		
		In 2008 the high luminosity flaring is seen at an average flux of $F=41$~mCrab. 
		Relating flux to mass accretion rate as $4\pi d^2 F = GM\dot{M}/R$, for a distance
		$d=3.5(1)$~kpc \citep{Galloway2006}, and bolometric correction
		factor of 2.14 \citep{Galloway2008}, the mass accretion rate was $3\E{-10}
		\Msol$~yr$^{-1}$ during the high luminosity flaring in the 2008 outburst. In 2011 the 
		high luminosity flaring appears at an average flux of $75$~mCrab, giving a mass 
		accretion rate of $5\E{-10}\Msol$~yr$^{-1}$. 
		For a canonical accreting neutron star ($M=1.4\Msol$, $R=10$~km and $B=10^8$~Gauss), we 
		find an $r_m$ of $\sim$21 km and $\sim$19 km for 2008 and 2011, respectively. 
		
		As a second estimate of the inner disk radius we can assume the upper kHz QPO frequency
		(see Table~\ref{tab:QPOParameters}) to be a proxy for the inner disk Kepler frequency.
		This way we obtain radii of 21--28 km. All these estimates are consistent with
		an inner disk cut off by the magnetosphere near the corotation radius.

	\subsection{Magnetic Reconnection}
		Magnetic reconnection has been proposed by \citet{Aly1990} as the origin of QPOs 
		that occur in the Rapid Burster at frequencies similar to those in SAX J1808 discussed
		here. In their magnetic reconnection model the stellar magnetic field threads the inner disk. 
		The differential rotation between the accretion disk and the magnetosphere shears the 
		field lines, building up magnetic energy. This energy is periodically released 
		in reconnection events that occur a few times per beat period between the inner edge of the accretion disk 
		and the neutron star spin. The reconnection events break up the accretion flow
		into blobs, which results in the observed flaring behavior.

		Because in this model the QPO frequency depends on the beat between the inner disk radius and
		the neutron star spin, it will be a function of mass accretion rate. An increasing mass 
		accretion rate pushes the disk inward. For $r_m > r_c$ this causes a decreasing QPO
		frequency, at $r_m = r_c$ the QPO frequency passes through zero, and for $r_m < r_c$
		the frequency will increase with mass accretion rate.
		For the high luminosity flaring we found $r_c \simeq r_m$, so the frequency is in
		the correct range. The higher frequency and mass accretion rate
		in 2011 with respect to 2008 then implies $r_m \lesssim r_c $, so that centrifugal
		inhibition is not an issue. However, the frequency and flux change between the outbursts,
		$\nu(F)$, is inconsistent with predictions. 
		
		The QPO amplitude predicted by \citet{Aly1990} for an inner disk radius inside $r_c$ 
		is only a few percent, whereas our observed amplitudes are at a few tens of percent.
		This instability is therefore unlikely to be the cause of the high luminosity flaring.

	\subsection{Interchange Instabilities}
		Interchange instabilities can occur at the boundary between the neutron star magnetosphere
		and the accretion disk. In the case of stable accretion, the accretion disk structure
		cannot be maintained within the magnetosphere. At the magnetospheric boundary 
		matter is forced to move along the magnetic field lines, forming an accretion funnel to the
		neutron star surface. An interchange instability can occur if it is energetically more
		favorable for the accreting matter to be inside the magnetosphere rather than outside 
		\citep{Arons1976}. Plasma screening currents allow some of the matter
		to slip between the magnetic field lines and form long narrow accretion streams directly
		to the neutron star equatorial plane.
		
		Numerical simulations of interchange instabilities in the context of accretion onto
		a weakly magnetized neutron star have been done by \citet{Romanova2008}. Such simulations confirmed
		the formation of equatorial accretion tongues and showed that they can co-exist with
		a stable accretion funnel \citep{Romanova2008, Kulkarni2008}. \citet{Kulkarni2008}
		further showed that a small number of accretion tongues can remain coherent for
		short periods of time, producing a QPO in the light curve.
	
		\citet{Kulkarni2009} found that accretion tongues create a QPO
		with the rotation frequency at the inner disk radius. The expected frequency
		of such a QPO is therefore much too high to explain the high luminosity flaring. 
		Alternatively, the frequency at which the tongues are formed and disappear may 
		also lead to a QPO. However, then the flux modulation of an individual 
		flare would not affect the mean flux and the pulsed emission similarly, rather
		the opposite might be expected. Since we observed the pulse amplitude to change 
		proportionally with flux variations due to the flaring, an interchange instability 
		origin can also be ruled out.
		
	\subsection{Unstable Dead Disk}
		When $r_m$ approaches $r_c$, the centrifugal force at the inner edge of the disk 
		can prevent accretion onto the neutron star \citep{Illarionov1975}. When the inner 
		disk radius remains smaller than $\sim1.3r_{c}$, the centrifugal force is not strong
		enough to accelerate matter beyond the escape velocity \citep{Spruit1993} and matter will accumulate
		at the magnetospheric radius \citep{Spruit1993, Rappaport2004}. The inner
		accretion disk can then be described with the dead-disk solution \citep{Sunyaev1977},
		which has been shown to be subject to an accretion instability 
		\citep{dAngelo2010}.
        
        The instability arises when the mass accretion onto the neutron star is suppressed
        and a mass reservoir builds up in the inner accretion disk. As the reservoir grows
        in mass it exerts more pressure on the magnetosphere, forcing the inner disk radius
        to move inwards. Once a critical radius is reached, the disk can overcome
        the centrifugal barrier and the reservoir empties in an episode
        of accretion.

        In the simplest form of this model, the inner edge of the accretion 
        disk oscillates near the corotation radius. While the reservoir builds up mass, the
		accretion onto the neutron star, and as such the pulsed emission, stops. Although 
		we find $r_m \sim r_c$, the second characteristic is in conflict with our observation
		of pulsations during the flux minima of the high luminosity flaring. 

        Recent investigations by \citet{dAngelo2010,dAngelo2012} show that 
        the range of mass accretion rates at which the instability can occur 
        is much larger than initially suggested by \citet{Spruit1993}. By 
        parameterizing the uncertainties in the disk/magnetosphere interaction 
        in terms of a length scale over which the accretion
		disk boundary moves during the instability, and a length scale 
		that gives the size of the disk/magnetic-field coupling region,
    	\citet{dAngelo2010}
		find that the Spruit-Taam instability can occur
        in two regions of parameter space. One region covers the original
        regime studied by \citet{Spruit1993}, while the other region, which they 
        call RII, extends to higher mass accretion rates. 
        This RII region was shown to occur together with continuous accretion
        \citep{dAngelo2012} and is a plausible candidate for the high luminosity
        flaring.

		The mass accretion rates explored by \citet{dAngelo2012} are
		parameterized in terms of a characteristic mass accretion rate
		\begin{align}
			\dot{m}_c \simeq 0.5 \E{-10} 
					&\left[ \frac{M}{1.4\Msol} \right]^{-5/3}
					\left[ \frac{B}{10^8 \mbox{ Gauss}} \right]^{2}
					\\ \nonumber
					&\times
					\left[ \frac{R}{10 \mbox{ km}} \right]^{6}
					\left[ \frac{\nu_s}{400 \mbox{ Hz}} \right]^{7/3}
					\Msol \mbox{ yr}^{-1}.
		\end{align}
		which is the mass accretion rate at which $r_m = r_c$.
		The instability is found to occur at mass accretion rates of 0.1--10$\dot{m}_c$, 
		with periods of 0.01--0.1$\tau_v$, where 
		\begin{align}
		\label{eq:viscous}
			\tau_{v} \simeq 40
				 	&\left[ \frac{\alpha}{0.1} \right]^{-4/5} 
					 \left[ \frac{\dot{M}}{4\E{-10} \Msol\mbox{yr}^{-1}} \right]^{-3/10} 
					\\ \nonumber
					&\times 
					\left[ \frac{M}{1.4 \Msol} \right]^{1/4} 
					\left[ \frac{R_{i}}{20 \mbox{ km}} \right]^{5/4} \mbox{ s},
		\end{align}
		is the viscous timescale at the inner edge of the accretion disk, $R_i$, with $\alpha$ the disk
		viscosity parameter. This means the instability can occur at the $3-5\E{-10} \Msol$yr$^{-1}$
		at which we observed the flaring, and the frequency
		of the instability is 0.25--2.5~Hz, which agrees well with the observed flaring frequency.

		The RII instability region is bounded by a lower and an upper mass accretion rate,
		which can explain why the high luminosity flaring is seen in a narrow flux window and
		why that window has the same width in 2008 and 2011. The difference in flux at which the window
		is located in the two flaring instances is more difficult to understand. Possibly this
		difference could be explained by a change in the relative size of the two 
		length scales governing the model, which is predicted to change the mean mass accretion rate of
		the instability window. This could also explain why the high luminosity
		flaring is not observed in 2002 and 2005 outbursts. However, why these length scales 
		would change between outbursts is unclear, especially as SAX J1808 is otherwise 
		very homogeneous in its outburst phenomenology.
			
		The dead-disk accretion instability can naturally explain both the high and low luminosity
		flaring if the latter is caused by the low mass accretion rate instability region as
		proposed by \citet{Patruno2009}. Owing to the lower luminosity, the flaring in the outburst 
		tail has $\tau_{v}\sim$60--100~s (Equation~(\ref{eq:viscous})),
		giving a frequency range of 0.1--1.7~Hz, which agrees with the observed low luminosity flaring
		frequency range. 		

\section{Conclusions}
\label{sec:Conclusions}
	We have reported on the discovery of a 1--5 Hz flaring phenomenon with $\sim$20\%--45\% 
	fractional rms appearing at the peak of the 2008 and 2011 outbursts of SAX J1808. 
	We have performed a detailed timing study of this high luminosity flaring and found that
	it is similar to the previously reported low luminosity flaring, which is observed in the prolonged outburst 
	tail. We found that pulse amplitude changes proportionally to flux variations in individual 
	flares, such that the pulse fractional
	amplitudes are the same within errors, implying that the flaring is most likely
	present in the accretion flow prior to matter entering the accretion funnel.
	
	We have considered multiple candidate mechanisms for the high luminosity flaring and find
	that the dead-disk accretion instability of \citet{dAngelo2012} provides the most
	plausible explanation. This model was previously proposed to explain the very similar 
	low luminosity flaring seen in other outbursts of SAX J1808. 
		
	If both the observed high and low luminosity flaring are indeed caused by the dead-disk 
	accretion instability like we suggest, then the observation of this type of variability 
	is a new observational indicator of a magnetosphere in the system, accessible over a wide 
	range of accretion rates. It would therefore be very interesting to search for this type 
	of flaring in other, non-pulsating, LMXB sources.

\acknowledgments
We would like to thank A. Patruno and D. Altamirano for discussions on the
data analysis and C. D'Angelo for discussions on the theoretical implications. 

\bibliographystyle{apj}

\end{document}